\documentclass[aps, nofootinbib, twocolumn,groupedaddress,floatfix,nofootnotesinbib,preprintnumbers]{revtex4}
\usepackage{graphicx}
\usepackage{epsfig}
\usepackage{epstopdf}
\usepackage{rotating}
\usepackage[mathscr]{eucal}
\usepackage{amsmath,amssymb,bm,graphics}
\usepackage[letterpaper=true]{hyperref}
\usepackage{color}

\let\mathbf=\bm

\newif\ifarxiv
\arxivfalse

\begin{document}
\def\Nfour{\mathcal N\,{=}\,4}
\def\Ntwo{\mathcal N\,{=}\,2}
\def\Nc{N_{\rm c}}
\def\Nf{N_{\rm f}}
\def\x{\mathbf x}
\def\q{\mathbf q}
\def\f{\mathbf f}
\def\v{\mathbf v}
\def\C{\mathcal C}
\def\w{\omega}
\def\vs{v_{\rm s}}
\def\S{\mathcal S}
\def\half{{\textstyle \frac 12}}
\def\twothirds{{\textstyle \frac 23}}
\def\third{{\textstyle \frac 13}}
\def\t{\mathbf{t}}
\def\T{\mathcal {T}}
\def\O{\mathcal{O}}
\def\E{\mathcal{E}}
\def\p{\mathcal{P}}
\def\H{\mathcal{H}}
\def\uh{u_h}
\def\R{\ell}
\def\Ro{\chi}
\def\del{\nabla}
\def\eps{\hat \epsilon}
\def\nn{\nonumber}
\def\K{\mathcal K}
\def\inf{\epsilon}
\def\cs{c_{\rm s}}
\def\A{\mathcal{A}}
\def\e{{e}}
\def\r{{\xi}}
\def\x{{\mathbf x}}
\def\w{{w}}
\def\rr{{\xi}}
\def\uo{{u_*}}
\def\u{{\mathcal U}}
\def\G{\mathcal{G}}
\def\Deltax{\Delta x_{\rm max}}
\def\L{{\bm L}}

\title
{Bremsstrahlung and black hole production from collisions of ultra-boosted particles at non-zero impact parameter}

\author{Y.~Constantinou$^1$\footnotemark, A~Taliotis$^{2}$\footnotemark}

\affiliation
   {Crete Center for Theoretical Physics, University of Crete$^{1,2}$, 71003 Heraklion, Greece and Theoretische Natuurkunde, Vrije Universiteit Brussel and The International Solvay Institutes Pleinlaan 2, B-1050 Brussels, Belgium$^2$}

\date{August, 2013}

\begin{abstract}

The collision of two massless, gravitationally interacting, point-like massless particles, boosted to the speed of light, colliding with an impact parameter b is being investigated. The collision takes place in four space-time flat dimensional background. A perturbative scheme is employed and the corrections to the energy momentum tensor and to the metric are computed and closed form formulas are provided. This includes the back-reaction on the metric after the collision. Including such corrections suggests that the tracelessness of the initial stress tensors of the colliding particles is preserved during and after the collision. The necessity for introducing an impact parameter in the perturbative treatment is highlighted and the breaking of the underlying perturbative approach at $b=0$ is motivated. In addition, the energy radiated in the form of gravitational bremsstrahlung radiation is discussed while an example from gravitational-waves collision is being studied.

\end{abstract}

\preprint{CCTP-2013-12}


\maketitle
\iftrue
\def\thefootnote{\fnsymbol{footnote}}

\footnotetext[1] {Email: \tt gkonst@physics.uoc.gr}
\footnotetext[2] {Email: \tt atalioti@vub.ac.be}

\def\thefootnote{\arabic{footnote}}
\fi

\parskip	2pt plus 1pt minus 1pt

\section{Introduction}

Gravitational ultra-relativistic collisions of two black holes, boosted to ultrarelativistic \cite{Aichelburg:1970dh} speeds in flat backgrounds have been discussed in several contexts, \cite{D'Eath:1976ri,D'Eath:1992hb,D'Eath:1992hd,D'Eath:1992qu,Amati:1993tb,Kohlprath:2002yh,Amati:2007ak,Veneziano:2008xa,Ciafaloni:2008dg,Eardley:2002re,Gal'tsov:2009zi,Sampaio:2013faa}. In fact, recently with the application of AdS/CFT in heavy ions, the interest in a numerical or an analytical approach to this problem has been growing rapidly \cite{Casalderrey-Solana:2013aba,Romatschke:2013re,DeWolfe:2013cua,Wu:2013qi,Wu:2012rib,Grumiller:2008va,Wu:2011yd} but in AdS backgrounds however.

The novel features of the present work are
\begin{itemize}

\vspace{-0.02in}
\item The usage of a different pertrurbative scheme\footnote{The expansion parameter will be defined below.} than the one employed in earlier works in the literature \cite{D'Eath:1976ri,D'Eath:1992hb,D'Eath:1992hd,D'Eath:1992qu,Coelho:2013zs,vanderSchee:2013pia,Coelho:2012sy,Coelho:2012sya,Herdeiro:2011ck,Sampaio:2013faa}.

\vspace{-0.05in}

\item The inclusion of a non-zero impact parameter $b$. 

\vspace{-0.05in}

\item The computation of the corrections to the energy-momentum tensor due to the back-reaction effects in the presence of $b$. In particular, the present approach allows us
to follow the collided particles either in the case where they will get trapped inside a horizon or if a horizon is not formed at all.
\vspace{-0.05in}

\item The first corrections of the metric including the back-reaction contribution of the above point. 

\vspace{-0.1in}

\end{itemize}

Summarizing our results we have:  
\begin{enumerate}
\item Derived a closed form formula for the first corrections of the metric and of the energy-momentum tensor\footnote{The corrections of the stress-tensor are quite general as they apply for any longitudinal profile of the colliding particles.}  in the presence of an impact parameter, including the back-reactions. 

\item Showed that the metric dependence on space and time is according to the ordering between the proper-time $\tau$ and the transverse distance $r=\sqrt{(x^1)^2+(x^2)^2}$ from the center of the shocks. In particular, according to fig. \ref{re}, in the $b=0$ limit there appears a $\tau \leftrightarrow r$ symmetry. A similar result has been observed in shock-wave collisions in AdS backgrounds applied to heavy ions \cite{Taliotis:2010pi}. Remarkably, the same observation has been made in \cite{Gubser:2010ze} using completely different methods and in particular a hydro-approach. The analogies between the problem studied in this work and heavy ions/Quark Gluon Plasma is further discussed in conclusion ii of sec. \ref{scon}.

\item Showed that for zero impact parameter, the perturbative approximation breaks down and there is an instantaneous and point-like violation of the conservation of the energy momentum tensor, which however, is hidden behind a horizon. 

\item Have highlighted the importance of introducing an impact parameter which regulates the produced radiation in the absence of any other transverse scale. 

\item Found that the total energy momentum tensor before the collision is traceless and remains traceless after the collision up to the order we have computed in our expansion. This could suggest that it is traceless to all orders and that tracelessness is conserved; a conjecture that is worth investigating further.

\item Calculated the energy emitted during a collision of gravitational waves and argued that the result is exact to all orders. 

\item Finally, we have examined the problem of shock-wave collision, in the spirit of the trapped surface analysis, produced by extended sources \cite{Taliotis:2012sx}. The results show that for dilute enough concentration of energy a black hole is not formed. This result seems as a manifestation of the Hoop Conjecture proposed by K. Thorn.

\end{enumerate}

This paper is organized as follows. 

In section \ref{1sw} we set-up the problem, specify the conventions and write the form of the metric that describes the superposition of the shockwaves before and after the collision.

In the next section \ref{B2B}, we compute the corrections to the energy-momentum tensor. These corrections are caused by the interaction of the shockwaves and are only present for positive times. We also discuss the region of applicability of our approach and we make connections with the time scales that a black-hole needs to be formed and equilibrate.

In section \ref{ctfe}, we show the tracelessness and conservation of the stress tensor, modulo an instantaneous-point-like violation of conservation for zero impact parameter which, is hidden behind a horizon. 

Sections \ref{FE} and \ref{CG} specify the gauge choice and state the field equations, up to second order taking into account the back-reactions found in section \ref{ctfe}.

Section \ref{SEq} deals with the solution of the field equations obtaining the second-order corrections to the metric.

A dimensional analysis argument is presented in section \ref{br} and yields the dependence of the total energy radiated in the form of gravitational waves from, the energy that is available and, the only other meaningful dimension-full parameter, the impact parameter. In fact, it is shown that for massless particles carrying positive definite energy, the perturbative scheme  that computes radiation fails. The origin of this result is identified and the total radiation for zero energy shock-waves is calculated.

We conclude, section \ref{scon}, with a summary and some comments of our main results.

Appendix \ref{A} contains a brief derivation/discussion of trapped surfaces formation from colliding extended matter distributions.

In Appendix \ref{B} we give the explicit forms of the polarization tensors.

\section{Setting up the problem} \label{Sup} 

\subsection{Single Shockwave Solution}\label{1sw}

We choose light cone coordinates for the longitudinal direction as these are the most natural for the problem at hand. They are defined by

\begin{equation}\label{LC}
x^{\mu}=(x^+,x^-,x^1,x^2) \hspace{0.3in}x^{\pm}=\frac{x^0\pm x^3}{\sqrt{2}}
\end{equation}
where $x^0$ is the time axis and $x^1,x^2,x^3$ cover $\mathbb{R}^{3}$. We use the mostly plus convention for the metric and we raise and lower indices with the flat metric $\eta^{\mu \nu}$.

In order to set up the problem, we begin by writing the metric of a black hole which is boosted to the speed of light along the $x^3$ direction. This metric, the Aichelburg-Sexl solution \cite{Aichelburg:1970dh}, has the form

\begin{align}\label{ds1}
\nonumber
ds^2 &=& &g_{\mu \nu}dx^{\mu } dx^{\nu} \\&=& -&2 dx^+ \, dx^- + t_1
(x^+,x^1,x^2)  d x^{+ \, 2} + d x_\perp^2 \hspace{0.06in}  
\end{align}
where
\begin{align}\label{s1}
\nonumber
t_1&=-4\sqrt{2}E G \delta(x^+) \log(p r), \, \vec{r}=(x^1,x^2)\,,& \\  r&=\sqrt{(x^1)^2+(x^2)^2}.&
\end{align}
The transverse part of the metric is flat, $d x_\perp^2 = (d x^1 )^2 + (d x^2)^2$, while $\delta(x^+)$ is the Dirac delta function, $E$ is the energy of the shockwave, $G$ is the four-dimensional Newton's constant and $p$ serves as an IR cutoff as explained in \cite{Aichelburg:1970dh} and in section \ref{IRc} below. The ansatz (\ref{ds1}) reduces the generally non-linear field equations to a single linear equation. To see this we write Einstein's equations with out a cosmological constant in the form

\begin{align}\label{Ein}
R_{\mu \nu}  = \kappa^2
\left( T_{\mu \nu} - \frac{1}{2} \, g_{\mu \nu} \, T \right) \notag \\
T = T_{\mu}^{\mu} = \, T_{\mu \nu} \, g^{\mu \nu}\hspace{0.15in}\kappa^2=8\pi G \ ,
\end{align}

where $R^{\mu \nu}$ is the Ricci tensor and $T_{\mu \nu}$ is the Energy momentum tensor.

Substituting the ansatz (\ref{ds1}) in Einstein's equations yields

\begin{equation}\label{Rmn}
R_{\mu \nu}=\delta_{\mu +} \delta_{\nu +} \left(4\sqrt{2}\pi E G \delta(x^+)\delta^{(2)}(\vec{r})  \right).
\end{equation}
The presence of Kroenecker's delta, $ \delta_{\nu +}$, shows that the only non zero component of the Ricci tensor is $R_{++}$. The $T_{++}$ component on the right hand side of previous equation corresponds to a massless point-like particle moving along $x^+=0$ and so with the speed of light and, it is normalized such that the particle has energy $E$ (see also (\ref{T1})). Then $t_1$ of (\ref{ds1}) is specified by solving the linear differential equation,
\begin{align}\label{R++log}
R_{++}&=-\frac{1}{2}\nabla_{\perp}^2 t_1(x^+,x^1,x^2) 
\notag \\&=\kappa^2 T_{++}  =4\sqrt{2}\pi E G \delta(x^+)\delta^{(2)}(\vec{r})
\end{align}
where $\nabla_{\perp}^2$ is the Laplace operator in two dimensions. Making use of
\begin{equation}\label{dlog}
\nabla_{\perp}^2 \log (k r)=2 \pi\delta^{(2)}(\vec{r})
\end{equation}
one verifies that $t_1$ is given by equation (\ref{s1}).

In the next section we collide two such shockwaves with an impact parameter, $b$. We thus shift the origin along the $x^1$ axis and the stress-energy tensor is given by 

\begin{equation}\label{T1}
T_{\mu \nu}^{(1)}=\frac{\sqrt{2}}{2}\delta_{\mu +} \delta_{\nu +}\left(E\delta(x^+)\delta(x^1-b)\delta(x^2) \right).
\end{equation}

The energy momentum tensor here describes a massless shockwave traveling at the speed of light along the $x^-$ direction. One can check that this energy momentum tensor is covariantly conserved. We note that this quantity, as mentioned, has been normalized so as to satisfy $\int T_{++} d^3x=E$.

\begin{figure}
\centering
\includegraphics[scale=0.45]{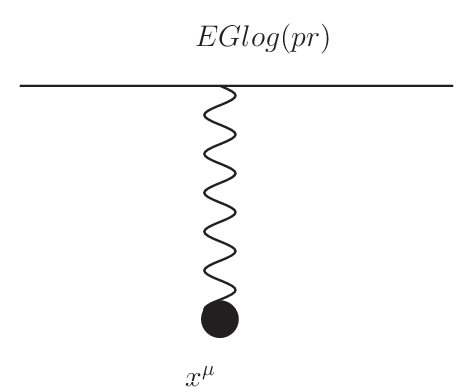}
\caption{A pictorial description of the shockwave solution, that is an exact solution to Einstein's equation. The bulk source, represented by the straight line, emits a graviton, represented by the curly line, with a coupling constant $E G \log(pr)$. The filed is measured at the point $x^\mu$.}
\label{vertex}
\end{figure}
Figure \ref{vertex} represents diagrammatically the configuration. It shows the gravitational field produced by a single graviton emission from the energy-momentum tensor of (\ref{T1}) with an effective coupling $E G \log(p r)$. The gravitational field is measured at the space-time point $x^\mu$. The sketch provides a suggestive diagrammatic expression for the metric of equation (\ref{ds1}).

\vspace{-0.1in}
\subsection{Superimposing two Shockwaves}\label{2sw}

 We assume that the background is a flat space-time with two shockwaves, $t_1$ and $t_2$ propagating in it. 
 
 In the language of the order counting, the background metric is described by terms of zeroth (flat piece of the metric) and of first order (the non interacting shocks) in the sources $t_1$ and $t_2$. The perturbations around this background, are considered second order in the sources and hence they involve products of $t_1 t_2$. The next orders in this expansion, which we will not compute, have the form $t_1^2 t_2$ and $t_1 t_2^2$ etc.

Our calculation is an expansion under the energy $E$ \footnote{Essentially the relevant dimensionless expansion parameter is $\frac{EG}{b}$ as section \ref{br} suggests.}. We use the superscripts $^{(n)}$ on the quantities $A^{(n)}$ to denote the n'th order of quantity $A$ in the given expansion. It is  evident that essentially this corresponds to the number of times the source $t_i$ ($i=1,2$) is inserted as a (right hand side) source for the propagator corresponding to the d'Alembert operator (see for example equation (\ref{deq})).

\begin{figure}
\centering
\includegraphics[scale=0.22]{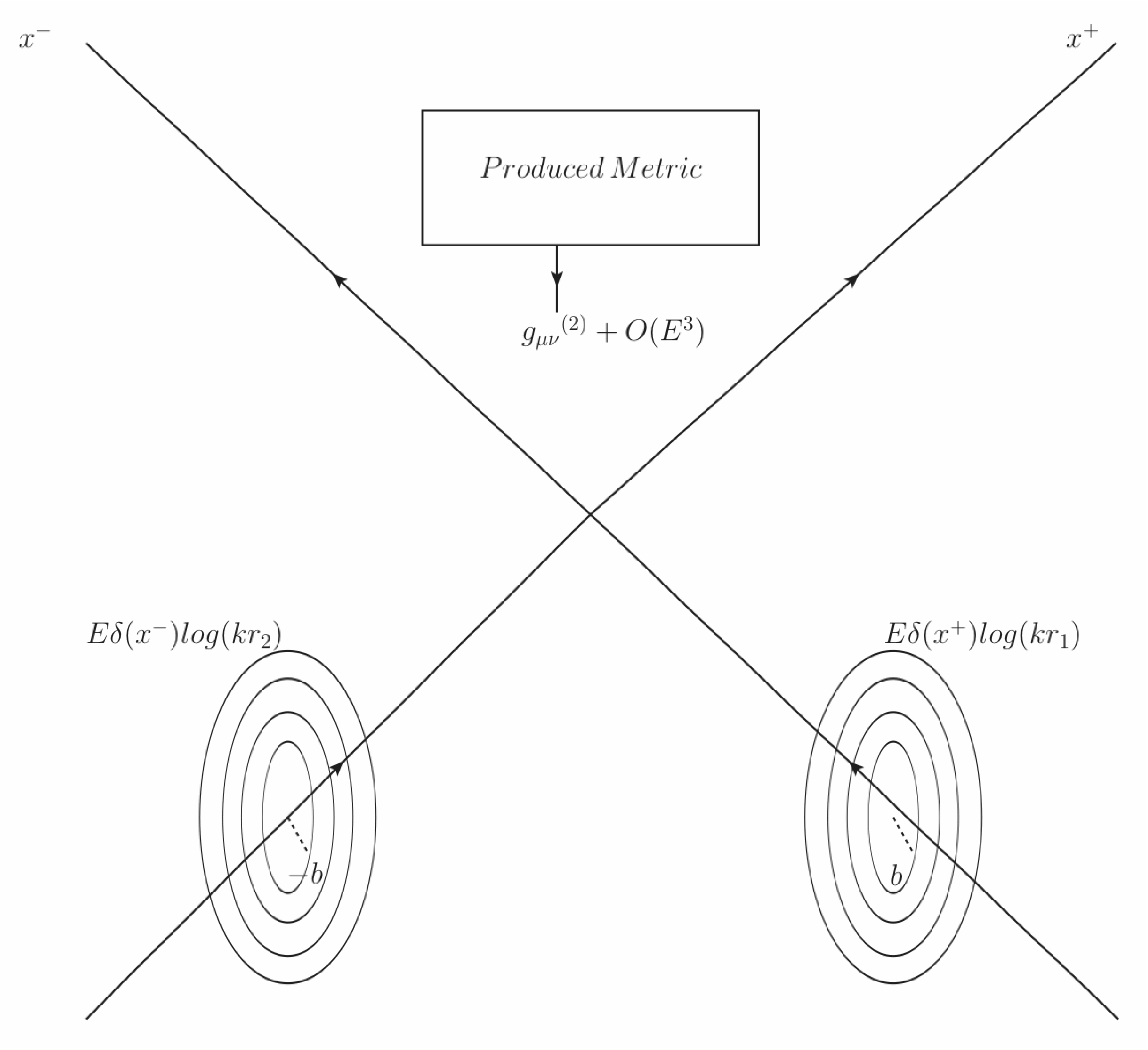}
 \caption{Presentation of the two shockwaves moving along the $x^{\pm}$ axis. Below the origin, time is negative and refers to pre-collision times. After the collision, the two shockwaves interact and produce a gravitational field in the forward light cone, described by the metric $g_{\mu \nu}^{(2)}$.}
  \label{offcenter}
 \end{figure}

In what follows we collide two such shockwaves moving at the speed of light towards each other with an impact parameter, $b$. After the collision we will compute their trajectories taking into account the back-reaction of the metric. The problem is treated classically. We must add to the energy momentum tensor of (\ref{T1}) a second part, describing the second shockwave,

\begin{align}\label{T2}
T^{(1)}_{--} \, = \frac{\sqrt{2} E}{2} \delta (x^-) \, \delta (x^1+b)\delta (x^2).
\end{align}
This collision is captured by fig. \ref{offcenter}, where  the two shockwaves are shown before the collision. Following \cite{Albacete:2009ji,Albacete:2008vs}, the metric that describes the process should look like
 
\begin{figure}
\centering
\includegraphics[scale=0.5]{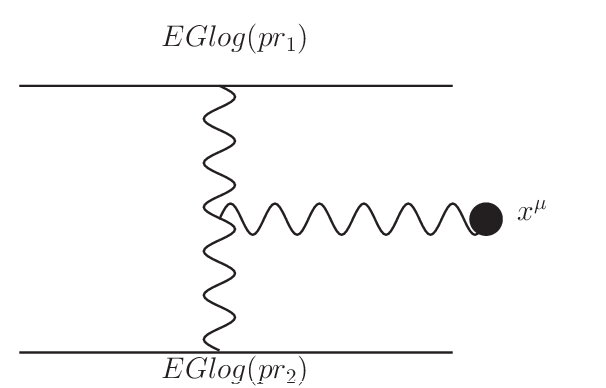}
\caption{Schematic representation of the $E^2$ corrections to the metric. Along with the diagrams of figure \ref{SelfInt4d}, the present diagram consists of the first non-trivial correction to the metric (\ref{s12}). The interaction of the metrics produced by each of the shockwaves is shown at the point of intersection of the graviton propagators while the gravitational field is measured at the space-time point $x^{\mu}$.}
\label{interaction}
\end{figure}

\begin{align}\label{s12}
ds^2 \, &= -2 \, dx^+ \, dx^- + d x_\perp^2 + t^{(1)}_1(x^+,x^1-b,x^2)  \, d x^{+ \, 2} \notag\\&
 + t^{(1)}_2(x^-,x^1+b,x^2)  \, d x^{- \, 2}  \notag\\&
+ \theta(x^+)\theta(x^-)g^{(2)} _{\mu \nu}(x^{\kappa},z)dx^{\mu}dx^{\nu}   + \ldots    , \notag\\&
\nonumber t^{(1)}_{1,2}(x^{\pm,}x^1 \mp b,x^2)& &\\
  &=-4 \sqrt{2}E G \log\left (k \sqrt{(x^1 \mp b)^2
 +(x^2)^2}\right) \delta(x^{\pm}).
\end{align}
A few explanations are in order: (a) The first two terms of this metric describe a flat Minkowski spacetime. 
The next two terms describe the two point particles that are moving on a collision course at the speed of light along the $x^3$ coordinate, with an impact parameter of $2b$, as shown in figure \ref{BTen}. These terms are of first order in $E$, since they describe the two shockwaves and not their interactions so far. Each of them is schematically described as a vertex diagram, shown in figure \ref{vertex}.  One can check that the metric, in this ansatz, satisfies the de Donder gauge up to first order in our counting.
(b) The next term, $g_{\mu \nu}^{(2)}$, is of second order in $E$ (as we will see, it appears as a product of $t_1 t_2$) and describes the interactions of the two pre-collision metrics. Essentially this term represents the superposition of the two vertices (as in figure \ref{interaction}) and only exists for times after the collision, a fact that is highlighted by the Heaviside theta function in (\ref{s12}). This is a consequence of retardation. For times before the collision where these corrections are zero, the remaining metric of the two shocks is an exact solution of the Einstein equations. (c) Our main result is the computation of the term that is quadratic in $E$, namely $g_{\mu \nu}^{(2)}$ which is presented schematically in fig. \ref{interaction} and \ref{SelfInt4d}.
(d) One might worry that in these coordinates the geodesics are discontinuous. As we will see in what follows, our approach applies for any profile of the shocks along $x^{\pm}$ and not only for $\delta(x^{\pm})$ profiles. Hence, we implicitly assume that we deal with regularized smooth functions\footnote{In any case, the physics should not depend on the coordinate system.}.
\section{Back-Reactions} \label{B2B} 

\subsection{Corrections to $T_{\mu \nu}$ and Geodesics}\label{Tgen}

The energy momentum tensor of equation (\ref{T1}), $T_{++}^{(1)}$, is conserved in the metric described by equations (\ref{ds1}) and (\ref{s1}). In fact, it is also conserved to all orders in $E$, even though in practice only terms linear in $E$ will appear in the equations. This has an intuitive explanation in the context of figure \ref{vertex}. Gravity is linear in view of the metric (\ref{ds1}) and the following equation
\begin{align}\label{con1}
\nabla^{\mu} T_{\mu \nu}=0
\end{align}
is true to all orders.

\begin{figure}
\centering
\includegraphics[scale=0.5]{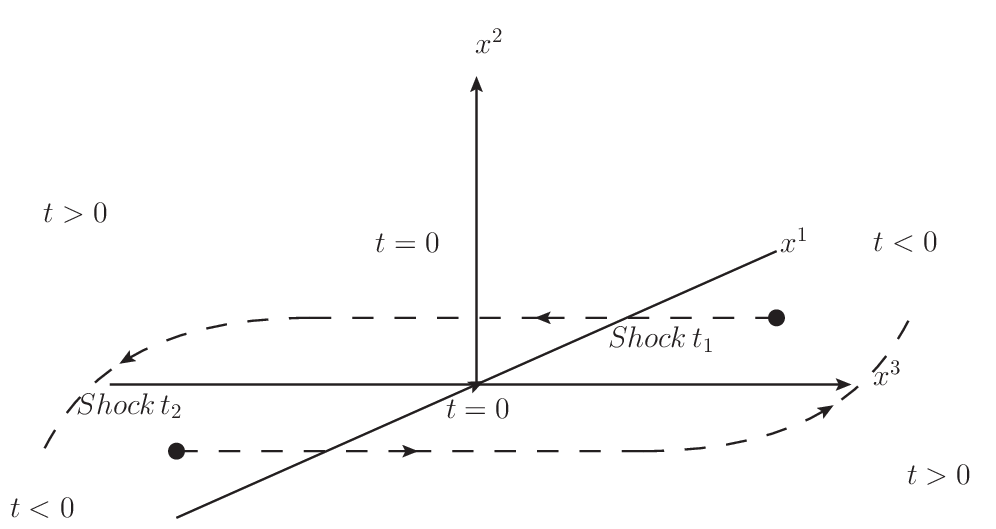}
\caption{The two shockwaves, represented as black dots, move on the trajectories shown by the dashed lines. Before the collision, they move along straight lines while after the collision, occurring at at t=0, their trajectories are modified and follow what is pictorially shown as a curved path. This results to a change in the $T_{\mu \nu}$.}
\label{BTen}
\end{figure}
\begin{figure}
\centering
\includegraphics[scale=0.5]{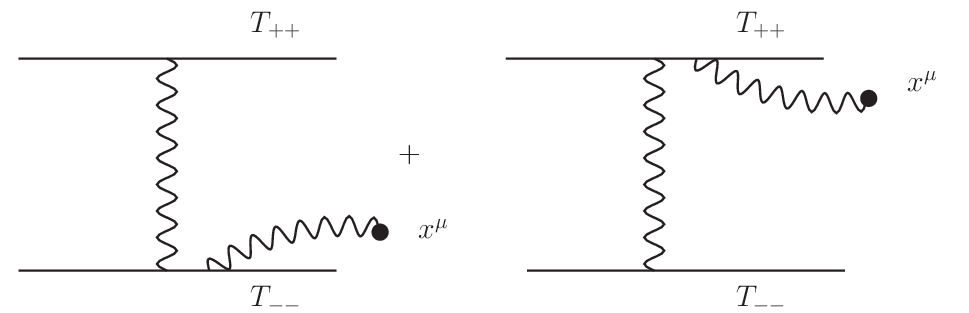}
\caption{The diagrams representing the backreactions. The emission of gravitons, curly lines, due to the self corrections to $T_{\mu \nu}$ are shown. Each shockwave moves in the gravitational field produced by the other. The gravitational field is measured at the point $x^\mu$.}
\label{SelfInt4d}
\end{figure}
 
It is also true that the combined energy momentum tensors $T_{++}^{(1)}$ and $T_{--}^{(1)}$ are conserved in the gravitational field of (\ref{s12}), when considering up to linear terms in $E$. This is an exact result for negative times. Once quadratic terms are considered, i.e. at positive times, the metric needs to be corrected, because the trajectory of the particles responsible for the shockwaves is altered due to their mutual interaction. In figure \ref{BTen} we show the trajectories of the two colliding shockwaves, while in figure \ref{SelfInt4d} we show the self-corrections of the two energy momentum tensors. This occurs because for positive times each shockwave is moving in the gravitational field produced by the other. As a cross check the total energy-momentum tensor needs to be conserved up to the given order in $E$ of the expansion.

\subsection{Calculating the Corrections for $T_{\mu \nu}$}\label{Tcor}
The two shockwaves are massless and are thus light-like. Papapetrou, \cite{Papa:1974}, has rigorously proven that for particles moving on null geodesics, the energy momentum tensor conservation is guaranteed. The energy momentum tensor is a sum of point-like stress-tensors (particles) and is given by

\begin{align}\label{Tpp}
T ^{\mu \nu}= \frac{\pi E G}{\kappa^2}  \sum_{(I=1)}^2   \dot{x}_{(I)}^{\mu}\dot{x}_{(I)}^{\nu}  \frac{1} {\sqrt{-g}}\delta^{(3)} \left(\vec{x}_{(I)}- \vec{x}_{(I)}(s_{(I)}) \right).         
\end{align}
The particle $I$ ($I=1,2$) entering (\ref{Tpp}) moves along the trajectory $ \vec{x}_{(I)}(s_{(I)})$, where the trajectory has been parametrized by the variable $s_{(I)}$. The quantity $\sqrt{-g}$ refers to the determinant of the total metric. The dots denote differentiation with respect to the variable $s(I)$. The detailed calculation of the back-reaction was presented in detail in \cite{Taliotis:2010pi}. Here we will only outline the method and quote the result.

This is a two step process. In  step (i) one should find the corrections to the geodesics of the one particle in the gravitational field of the other. In practice, the perturbative solution to the equations

\begin{align}\label{geo}
\ddot{x}^{\mu}_I+\Gamma_{J;\nu\rho}^{\mu}  \dot{x}_I^{\nu}\dot{x}_I^{\rho}=0 \hspace{0.3in}I,J=1,2 \hspace{0.3in}I\neq J
\end{align}
is required. These are interpreted as the motion of particle $I$ in the gravitational field of the particle $J$  (due to $\Gamma_{J;\nu\rho}^{\mu}$ where $\Gamma$ are the Christoffel symbols) and vice versa; this is precisely the meaning of the subscripts $I$ and $J$. Step (ii) makes use of the result of step (i) by substituting  the corrections of the initial trajectories $x^{\mu}$ from (\ref{geo}) inside (\ref{Tpp}). Expanding all the functions to the order that is consistent with the perturbation yields the result.

\underline{Second order corrections to the total $T_{\mu \nu}$}
\vspace{0.15in}

We present here the quadratic (in $E$) corrections to the total energy momentum tensor:
\begin{subequations}\label{Tmn2}
 \begin{align}
&  (T_{+-})^{(2)}= -  (T^{+-})^{(2)}=    \frac{1}{4}\frac{ 1}{ \kappa^2}     \left(   t_{2} \nabla_{\bot}^2 t_{1} + t_{1} \nabla_{\bot}^2 t_{2} \right) \label{T+-}   \\&
(T_{++})^{(2)}= \frac{1}{4}\frac{ 1}{\kappa^2} \int  dx^-\Big(  t_{2}  \nabla_{\bot}^2 t_{1,x^+}\notag\\& +\nabla_{\bot}^2 t_{1,x^1} \int dx^- t_{2,x^1}+\nabla_{\bot}^2 t_{1,x^2} \int dx^- t_{2,x^2}\Big)\label{T++} \\&
(T_{--})^{(2)}= \frac{1}{4}\frac{ 1}{\kappa^2} \int  dx^+\Big(  t_{1} \nabla_{\bot}^2 t_{2,x^-} \notag\\& +\nabla_{\bot}^2 t_{2,x^2} \int dx^+ t_{1,x^2}+\nabla_{\bot}^2 t_{2,x^1} \int dx^+ t_{1,x^1}\Big)\label{T--}\\&
    (T_{+1})^{(2)}=              \frac{1}{4}\frac{1}{\kappa^2} \nabla_{\bot}^2 t_{1} \int  dx^- t_{2,x^1}
\notag\\&
   (T_{-1})^{(2)}=              \frac{1}{4}\frac{ 1}{\kappa^2} \nabla_{\bot}^2 t_{2} \int  dx^+ t_{1,x^1}         \label{T+-1}\\ &
    (T_{+2})^{(2)} = 	        \frac{1}{4}\frac{ 1}{\kappa^2} \nabla_{\bot}^2 t_{1} \int  dx^- t_{2,x^2}
\notag\\&
 (T_{-2})^{(2)} = 	        \frac{1}{4}\frac{ 1}{\kappa^2} \nabla_{\bot}^2 t_{2} \int  dx^+ t_{1,x^2}\label{T+-2}\\&
 (T_{11})^{(2)}= (T_{22})^{(2)}= (T_{12})^{(2)}=0.
\end{align}
\end{subequations}

It will prove to be useful to rewrite the first order energy momentum tensors of (\ref{T1}) and (\ref{T2}) using (\ref{dlog}) yielding 
\begin{align}\label{T12nab}
T_{++}^{(1)}=\frac{1}{2\kappa^2} \nabla_{\bot}^2 t_1, \hspace{0.3in} T_{--}^{(1)}=\frac{1}{2\kappa^2} \nabla_{\bot}^2 t_2.
\end{align}
The expressions that will be most useful for the following analysis will be the compact expressions of equations (\ref{Tmn2}). Nevertheless we also wish to show, for completeness, the energy momentum tensor explicitly in terms of the coordinates. Defining 
\begin{align}\label{r12}
 \vec{r_1}=   \vec{r}- \vec{b_1}               \hspace{0.3in}            \vec{r_2}=   \vec{r}- \vec{b_2} .
\end{align}
and substituting (\ref{dlog}) in  (\ref{Tmn2}) we obtain

\begin{subequations}\label{Tmnc}
 \begin{align}
 \nonumber
&(T_{+-})^{(2)}=    \frac{16 \pi E^2 G^2}{\kappa^2}\log(2p|b|) \delta(x^+)\delta(x^-) \\&
\times  \left( \delta^{(2)}(\vec{r_1})  +   \delta^{(2)}(\vec{r_2})) \right) \label{T+-c}   \\&
(T_{++})^{(2)}=  \frac{16 \pi E^2 G^2}{\kappa^2} \theta(x^-)\Bigg[\log(2p|b|) \delta'(x^+)\delta^{(2)}(\vec{r_1})\notag\\& \hspace{0.55in}+\frac{x^-}{x^1+b}\delta(x^+)\delta'(x^1-b)\delta(x^2) \notag\\&
\hspace{0.55in}+\frac{x^- x^2}{4b^2+(x^2)^2}\delta(x^+)\delta(x^1-b)\delta'(x^2) \Bigg] \label{T++c}\\&
    (T_{+1})^{(2)}= \frac{8\pi E^2 G^2}{\kappa^2 |b|} \theta(x^-)\delta(x^+)\delta^{(2)}(\vec{r_1})   \label{T+-1c}\\ &
    (T_{+2})^{(2)} = 	     0. \label{T+2c}
\end{align}
\end{subequations}
The asymmetry between $T_{+1}^{(2)}$ and $T_{+2}^{(2)}$ is due to the fact that the impact parameter $\vec{b}$ has only an $x^1$ component by assumption. The other components of the energy momentum tensor are completely symmetric and are trivially obtained by exchanging simultaneously $+ \leftrightarrow-$ and $b \leftrightarrow-b$ respectively.

The important feature of (\ref{Tmn2}) is that all the corrections for $T_{\mu \nu}$ involve either a product of three delta and one theta functions or a product of four delta functions. This implies that equation (\ref{Tmnc}) provides localized corrections that either apply only in the forward light cone or corrections that apply only at one point and only at one instant. Essentially this expresses the fact that the geodesics are discontinuous and as a result $T_{\mu\nu}$ experiences sudden changes as is to be expected. Using regularized functions would smooth out the underlying sudden ``kicks" and it would introduce a finite interaction time.

\subsection{Region of validity and the physical meaning of the IR cut-off}\label{IRc}

I. $b\rightarrow \infty$: One would expect that in this limit the particles would not interact. However, the single shock metric behaves as $\log(p r)$ (see (\ref{ds1}) and (\ref{s1})) where $p$ is an IR cut-off resulting from the Aichelburg- Sexl ultrarelativistic boost\footnote{Where in one of the intermediate steps of this large boost one has to integrate $1/\sqrt{r^2+x_3^3}$ along the boost, that is along $x_3$, from $-\infty$ to $+\infty$. This obviously diverges logarithmically and an IR cut-off is placed at large $x_3=1/p$.}. Such a logarithmically large field at large distances away from the source creating it implies that a second source located very far from the first one, will feel that large field and vice versa\footnote{Such an ambiguity arises as a result of the superposition of the two colliding metrics. One can check that the Riemann tensor of the single shock tends to zero as $r \rightarrow 0$. However, once the two metrics of the two shocks are superimposed, in the forward light-cone, this is no longer the case.}. This is clearly problematic. Therefore, combining this and the fact that $p$ is an IR cut-off, we argue that the metric (\ref{ds1}) and as a consequence the metric (\ref{s12}) makes sense for distances less than $\sim 1/p$. Thus, trusting the solution up to transverse distances of the order of $1/p$ we deduce that large $b$ implies $pb\sim 1$ in which case $T_{\mu \nu}^{(2)}\rightarrow 0$ as should. On the other hand, the current set-up may be treated perturbatively when the inequality $EG/b\ll1$ is valid  (see section \ref{br}).  This inequality can apply simultaneously with the inequality $b\ll1/p$. Hence, our set-up is justified when
\begin{align}\label{Ebr}
EG\ll b\ll1/p\equiv r_{IR}\,\,\, \mbox{and}\,\,\, r\in(0,1/p).
\end{align}
We will see this necessity of placing an IR cut-off in more detail in section \ref{br}. It will become evident that in higher dimensions such a cut-off is not required and that this is an artifact of 4-dimensions.

Equation (\ref{Tmnc}) suggests that there is another kinematical restriction for the applicability of our perturbative treatment. In particular the $T_{\pm \pm}^{(2)}$ components grow for large $x^{\pm}$. Obviously, we should expect that the corrections to $T_{\mu \nu}$ could not grow infinite with time. Hence, restricting the expansion to lower orders is consistent if we restrict the kinematical region where we trust our result, that is from $x^\pm=-\infty$ up to $x^\pm \sim EG$.
 
Therefore, our expansion is an early times expansion close to the collision point $x^3=0$.
It is expected that higher graviton exchanges, than those appearing in fig. \ref{interaction} and \ref{SelfInt4d}, will unitarize the corrections to $T_{\mu \nu}$ as $x^{\pm} \rightarrow \infty$. In fact, the same restrictions on $x^{\pm}$ apply according to earlier works in the literature about gravitational-wave collisions in AdS$_5$ backgrounds \cite{Albacete:2009ji,Albacete:2008vs}. There, the set-up is similar to the one of this work. In fact, through resummations of multiple graviton exchanges in \cite{Albacete:2009ji} , it is found that the shocks eventually will decay at large $x^{\pm}$ at times scales set by the energy of the shocks. Furthermore, it is found that the same time scale where the shocks stop sets also the thermalization time \cite{Kovchegov:2009du}\footnote{Where thermalization time is estimated as $\sim E^{1/3}$ for these AdS$_5$ backgrounds.} . By the same analogy, one would expect that a similar result would apply here. That at times of the order of $EG$ the remaining shocks that continue to move on the light-cone will be completely wiped out while a black hole will be formed\footnote{In this part of the discussion we assume assume $b\ll EG$ in which case the problem can only be studied numerically unless some resummations techniques, along the lines of  \cite{Albacete:2009ji}, can be engineered.}. Certainly, the final word on the metric evolution belongs to the numerical relativity community.

II. $b\rightarrow 0$: It seems that $T_{\mu \nu}^{(2)}$ diverges. As we will see, this is one of the many manifestations of the same fact: the problematic behaviour of the perturbative treatment in this limit.

\section{Conservation, Tracelessness and Field Equations}\label{ctfe}

We have so far computed the energy momentum tensor up to second order in $E$. Calculating the divergence of $T_{\mu \nu}$ up to second order in $E$, one finds
\begin{align}\label{con}
& \left((\nabla^{\mu})^{(0)}+ (\nabla^{\mu})^{(1)}\right)   \left( (T_{\mu\nu})^{(1)}+(T_{\mu\nu})^{(2)} \right) =\notag\\&
\delta_{\pm \nu} \nabla_{\bot}^2t_1\nabla_{\bot}^2t_2+O(E^3)\notag\\&
  \sim \delta_{\pm \nu}\delta^{(2)}(r)\delta^{(2)}(b)\delta(x^+)\delta(x^-)+O(E^3).
\end{align}
The operator $\nabla$ denotes the covariant derivative. In arriving to the second line of (\ref{con}) it is not necessary to substitute the precise profiles of $t_1$ and $t_2$. Hence, this suggests that (\ref{Tmn2}) has a rather general applicability. On the other hand, in arriving to the third line of (\ref{con}), which shows conservation for non-zero impact parameter $b$, we had to use equation (\ref{dlog}). We thus conclude that equation (\ref{Tmn2}) is consistent with conservation and hence correct, for any longitudinal profile\footnote{Of $T_{\mu\nu}$ of the initial particles as a function of $x^{+}$ and $x^{-}$; not just for the $\delta(x^{\pm})$ profiles.} provided that their transverse profile is localized and separated. To this end, one could argue that for zero impact parameter there exists an instantaneous violation of the conservation of $T_{\mu\nu}$ and hence of the Bianchi identities at the point $x^1=x^2=x^+=x^-=0$. However, on one hand, according to section \ref{br}, our perturbative treatment breaks down when $b=0$ and on the other hand, as we know from \cite{Eardley:2002re}, a black hole will be formed and the violation will be hidden behind the horizon.

One could imagine smoothening out the transverse distribution of the stress tensor of the initial point-like particles. For instance, one could use Gaussians instead of delta functions. Then at a first glance, the right hand side of equation (\ref{con}) seems to be a product of Gaussians implying an instantaneous but non-localized violation of conservation. This would raise doubts about the validity of our calculation\footnote{We thank S. Gubser and G. Horowitz for related discussions.}. However, our present derivation for the corrections due to back reactions of $T_{\mu \nu}^{(2)}$ given by equation (\ref{Tmn2}), is based on equation (\ref{Tpp}) which explicitly assumes point-like sources. In other words (\ref{Tmn2}) is not applicable for extended sources and the derivation in such a case needs to be modified. Therefore, it is not a surprise that equation (\ref{Tmn2}) for extended sources would invalidate conservation everywhere in space-time. Certainly, considering extended sources is an interesting generalization which we postpone for a future investigation. To this end, in appendix \ref{A} we show, using a trapped surface criterion, that when the transverse distribution is dilute enough, a black-hole can not be formed \cite{Taliotis:2012sx}. This implies that under some circumstances, a horizon will not be formed and hence the non-local violation of conservation will not be hidden. Evidently, this is another indication that equation (\ref{Tmn2}), despite its invariant-looking form, needs to be modified for extended sources along the transverse directions.

Calculating the trace of the stress-tensor to second order yields
\begin{align}\label{Tr}
\nonumber T &=& &g^{\mu \nu}T_{\mu \nu}&\\
&=& &(g^{\mu \nu})^{(1)}(T_{\mu \nu})^{(1)}+(g^{\mu \nu})^{(0)}(T_{\mu \nu})^{(2)}=O(E^3)&
\end{align}
up to quadratic order in $E$. This simplifies Einstein's equations as follows
\begin{align}\label{EE}
R_{\mu \nu}  = \kappa^2
T_{\mu \nu}+O(E^3)  \hspace{0.35in}\kappa^2=8\pi G.
\end{align}
Equation (\ref{Tr}) shows that the stress-tensor due to the "cross-talk" between the stress-tensor corresponding to the shock $t_1$ and the stress-tensor corresponding to the shock $t_2$, is precisely cancelled by the corrections due to the back-reaction contribution. As a result, we started with an energy momentum tensor that was traceless for negative times and found that, up to second order, it is also traceless for positive times. This suggests that the energy momentum tensor could be  traceless to all orders and all times, implying a conservation of tracelessness conjecture. This is worth investigating further.
 
\section{Field Equations} \label{FE} 

\subsection{ Field Equations to $O(E^2)$ }\label{FE1}
We now proceed to explicitly construct the field equations of (\ref{EE}) up to second order. The zeroth order terms satisfy the Einstein equations trivially since $R_{\mu \nu}^{(0)}=T_{\mu \nu}^{(0)}=0$. The first order terms $R_{\mu \nu}^{(1)}$ from equation (\ref{s12}) satisfy the field equations with a right hand side given by $T_{\mu \nu}^{(1)}$, equations (\ref{T1}) and (\ref{T2}). 
Hence, we only require the second order terms to satisfy
\begin{align}\label{Rmn2}
R_{\mu \nu}^{(2)} = \kappa^2T_{\mu \nu}^{(2)}
\end{align}
where the second order energy momentum tensor, $T_{\mu \nu}^{(2)}$, is given by (\ref{Tmn2}) and it corresponds to the diagram of figure \ref{SelfInt4d}. We split the second order Ricci tensor in two parts. First, a known part that is due to the two shockwaves and is a product of the two vertices (see figure \ref{vertex}) of $t_1$ and $t_2$. We will use the notation $-S_{\mu\nu}$ for this part and it corresponds to the diagram of figure \ref{interaction}. The second part comes from the quadratic terms in $E$ of the metric, $g_{\mu \nu}^{(2)}$ and we will use the notation $(R_{\mu \nu}^{(2)})_g$. We now proceed to expand (\ref{Rmn2}) up to quadratic order in $E$ thus obtaining
\begin{align}\label{Rmn2gt}
(R_{\mu \nu}^{(2)})_g-S_{\mu \nu}^{(2)} = \kappa^2T_{\mu \nu}^{(2)}.
\end{align}
A more suggestive way of writing this expression is
\begin{align}\label{Rmn2gtT}
(R_{\mu \nu}^{(2)})_g = \kappa^2T_{\mu \nu}^{(2)}+S_{\mu \nu}^{(2)}\ ,
\end{align}
where $S_{\mu \nu}^{(2)}$ is considered to be an effective energy momentum tensor, contributing to the total one. All terms in the right hand side of this equation are known. To simplify the notation we will be suppressing the superscripts denoting the order of the terms. We will restore them wherever it is necessary.

\section{Choosing the Gauge and Field Equations}\label{CG}
In this section we specify the gauge choice and present the field equations including the back-reacted contribution found in section \ref{Tcor}. Working in the harmonic (de Donder) gauge
\begin{align}\label{gc}
g_{\mu \nu},^{\mu}-\frac{1}{2}{g^\mu}_\mu,_{\nu}=0
\end{align}
the Einstein's equations (\ref{Rmn2gtT}) in component form read
\begin{subequations}\label{deq}
\begin{align}
(++)\hspace{0.4in}& \Box g_{++}=- \frac{1}{2}  \int  dx^-\Bigg(  t_{2}  \nabla_{\bot}^2 t_{1,x^+} +\nabla_{\bot}^2 t_{1,x^1} \notag\\&
\times  \int dx^- t_{2,x^1}+\nabla_{\bot}^2 t_{1,x^2} \int dx^- t_{2,x^2}\Bigg),\label{++}\\
(+-)\hspace{0.4in}&  \Box g_{+-}=- \frac{1}{2}\left(   t_{2} \nabla_{\bot}^2 t_{1} + t_{2} \nabla_{\bot}^2 t_{1} \right)-t_{1,x^1}t_{2,x^1}\notag\\&
-t_{1,x^2}t_{2,x^2}+\frac{1}{2}t_{1,x^+}t_{2,x^-},\label{+-}\\
(+1)\hspace{0.46in}& \Box g_{+1}=-\frac{1}{2} \nabla_{\bot}^2 t_{1} \int  dx^- t_{2,x^1}+\frac{1}{2}t_{1,x^+}t_{2,x^1},\label{+1}\\
(11)\hspace{0.5in}& \Box g_{11}=t_{1,x^1}t_{2,x^1} + t_{1,x^1x^1}t_{2} + t_{1}t_{2,x^1x^1}=0,\label{11}\\
(12)\hspace{0.5in}  & \Box g_{12}=
\frac{1}{2} t_{2,x^2} t_{1,x^1}+ \frac{1}{2}  t_{1,x^2}t_{2,x^1}+t_{2}t_{1,x^1x^2} \notag\\&
+t_{1}t_{2,x^1x^2} =0.\label{12}
\end{align}
\end{subequations}
The integration limits have been suppressed, as will be done in the rest of this paper. For instance $\int t_{2,x^1} dx^+$ implies $\int_{-\infty}^{x^+} \partial_{1} t_2(x'^+,x^1,x^2) dx'^+$ etc. The operator $\Box$ denotes the d'Alembert operator in flat space-time, i.e.
\begin{align}\label{box}
\Box \equiv \eta^{\mu \nu}\partial_{\mu} \partial _{\nu}=-2 \partial_{x^+}\partial_{x^-}+\nabla_{\bot}^2.
\end{align}
In order to obtain equation (\ref{deq}), equation (\ref{Tmn2}) was employed. We now have a set of differential equations, equations (\ref{deq}), which, we will proceed to solve in the next section utilizing the appropriate boundary conditions.
\section{Solving the Field Equations and Causality} \label{SEq} 

\subsection{Green's Function and Boundary Conditions}\label{seq1}
We look for causal solutions of (\ref{deq}). The d'Alembert operator has a known retarded Green's function in light-cone coordinates

\begin{align}\label{GF}
G(x^{\mu} - x'^{\mu})=-\frac{1}{4\pi}\frac{\theta(x^+-x'^+)\theta(x^--x'^-)}{\frac{1}{\sqrt{2}} \left((x^+-x'^+)+(x^--x'^-)\right)} \notag\\
\times\delta \left(\sqrt{2(x^+-x'^+)(x^--x'^-)}-|\vec{r}-\vec{r'}| \right)\ ,
\end{align}
where $\vec{r}=(x^1,x^2)$ as defined in (\ref{s1}), $\theta$ is the Heaviside theta function. The Green's function satisfies

\begin{align}\label{boxGF}
\Box G(x^{\mu} - x'^{\mu})=\delta(x^+-x'^+)\delta(x^--x'^-)\delta^{(2)}(\vec{r}-\vec{r'}).
\end{align}

\subsection{Integrations }\label{LCP}

As usual, we integrate the product of the right hand side of (\ref{deq}), which plays the role of the source with (\ref{GF}) over all space-time. It will be useful to define the following vectors
\begin{align}\label{b}
\vec{b_1}=(b_{11},b_{12}) \hspace{0.3in}\vec{b_2}=(b_{21},b_{22}). 
\end{align}

In the right hand side of (\ref{deq}) terms of the form $\partial_{{x^i}_a}\left( t_1 t_2 \right)$ and $\partial^2_{{x^i}_a{x^j}_c}\left( t_1 t_2 \right)$ appear. The subscript ($a$) refers to the source (taking the value 1 or 2) and the superscripts refer to the transverse coordinate (i) with respect to which the source is being differentiated (also taking values 1 and 2). It is simpler to exchange the differentiations over the spatial coordinates $x_{1,2}$ with derivatives over the vectors $b_{1,2}$, $\partial_{x_a^i} \rightarrow -\partial_{b_{a i}}$. At the end of the calculation the limits 
\begin{align}\label{lim}
 \vec{b_1} \rightarrow  (b,0) \hspace{0.3in}  \vec{b_2}\rightarrow  (-b,0)
\end{align}
must be taken.

Equations (\ref{++})-(\ref{12}) involve the product $t_1t_2$ differentiated with respect to the transverse coordinates. We have already exchanged these differentiations with derivatives with respect to $b$'s. As a result, one can verify that the transverse convolution of the sources with the Green's function, involves the following integral 

\begin{align}\label{Jin}
{\cal J} (r_1,r_2,\tau)=\frac{1}{2 \pi \tau}\int_{0}^{\infty}\int_{0}^{2\pi}r'dr'd\phi' \delta(\tau-r') \notag\\
\times \log(p |\vec{r'}+\vec{r_1}|) \log(p |\vec{r'}+\vec{r_2}|)\notag\\
=\int  \frac{d^2 qd^2l}{(2 \pi)^2}  \frac{e^{i\vec{q} \vec{r_1}} e^{i\vec{l} \vec{r_2}}}{q^2l^2} J_0\left(  \tau |\vec{l}+\vec{q}|      \right)
\end{align}
where we have introduced a new parameter, the proper time $\tau$, defined as $\tau=\sqrt{2 x^+ x^-}$. The second equality comes from expanding the logarithms in Fourier space and performing the spatial integrations. Both of the momentum integrations have been performed in \cite{Taliotis:2010pi} and the  result reads
\begin{align} \label{J}
{\cal J} (r_1,r_2,\tau)&=\theta(r_1-\tau)\theta(r_2-\tau){\cal J}_1 (r_1,r_2,\tau)\notag\\&
+ \theta(\tau-r_2)\theta(r_1-\tau){\cal J}_2 (r_1,r_2,\tau) \notag\\&
+ \theta(\tau-r_1)\theta(r_2-\tau){\cal J}_3 (r_1,r_2,\tau)\notag\\&
+\theta(\tau-r_1)\theta(\tau-r_1){\cal J}_4 (r_1,r_2,\tau)
\end{align}
where the ${\cal J}$'s are defined with the help of table \ref{ta1} and the expression 
\begin{subequations}\label{Ja}
\begin{align}
{\cal J}&(\tau,r_1,r_2) \equiv \ln(\xi_> k)\ln(\eta_> k)+\frac{1}{4}\Big[ Li_2\left( e^{i\alpha}\frac{\xi_<\eta_<}{\xi_>\eta_>} \right) 
\notag\\&
\hspace{0.6in}+ Li_2\left( e^{-i\alpha}\frac{\xi_<\eta_<}{\xi_>\eta_>} \right)\Big], \label{JJ}\\
&\xi_{>(<)}=max(min)(r_1,\tau) \hspace{0.1in} \eta_{>(<)}=max(min)(r_2,\tau), \label{ke}\\
&\hspace{1in}\vec{r_1}.\vec{r_2}= r_1 r_2 \cos\alpha. \label{a} 
\end{align}
\end{subequations}

We denote the angle between $\vec{r}_1$ and $\vec{r}_2$ with $\alpha$ while $Li_2$ is the dilogarithm function. Notice that ${\cal J}$ is real as expected. 

The value of ${\cal J}$ appears to depend on the ordering of $r_1$, $r_2$ and $\tau$. There are six independent ways of ordering the three variables. One may observe that when $r_1,r_2>\tau$ or/and $r_1,r_2<\tau$ applies yields the same value for $J$ regardless of the ordering of $r_1$ and $r_2$. This reduces the total number of independent orderings to four which are summarized in table \ref{ta1} \footnote{${\cal J}_2$ for example is ${\cal J}(\xi_>\eta_>=\tau r_1,\xi_<\eta_<=\tau r_2)$ where ${\cal J}(\xi_>\eta_>,\xi_<\eta_<)$ is given in equation (\ref{Ja}). }. In particular, in the $b=0$ limit there exist only two orderings $\tau>r$ and $\tau<r$ suggesting a $\tau \leftrightarrow r$ symmetry\footnote{In this case where $b=0$ all order terms are required for calculating some quantities such as the radiation (see section \ref{br}). However, the argument for the $\tau \leftrightarrow r$ symmetry still applies.}.

The integrations over $x^{\pm}$ are trivial since they involve Dirac delta functions. We now present all the components of the metric, and refer the reader to \cite{Taliotis:2010pi} for the details of the calculation.

\begin{table}
\caption{We defined $\vec{r}_{1,2}=\vec{r}-\vec{b}_{1,2}$}
\centering
\begin{tabular}{c|cccc|c|c}
\hline\hline
cases & $\xi_>$ & $\eta_>$ & $\xi_>\eta_>$ & $\xi_<\eta_<$ & ${\cal J}_i$ & region (see figure \ref{re})\\
\hline\hline
1 & $r_1$ & $r_2$ & $r_1r_2$ & $\tau^2$ & ${\cal J}_1$ & I\\
2 & $r_1$ & $\tau$ & $\tau r_1$ & $\tau r_2$ & ${\cal J}_2$ & II$^{\prime}$\\
3 & $\tau$ & $r_2$ & $\tau r_2$ & $\tau r_1$ & ${\cal J}_3$ &II\\
4 & $\tau$ & $\tau$ & $\tau^2$ & $r_1r_2$ & $ {\cal J}_4$ &III\\
\hline\hline
\end{tabular}
\label{ta1}
\end{table}
 
\underline{The Formula for $g_{\mu \nu}^{(2)}$}

Using the compact notation $\lim_{\vec{b}_{1,2}\to (\pm b,0)} \equiv \lim_{\vec{b}_{2}\to (- b,0)} \lim_{\vec{b}_{1}\to ( b,0)}$ we finally have

\allowdisplaybreaks
\begin{subequations}\label{gmn2}
\begin{align}
 g_{++}^{(2)} &=\lim_{\vec{b}_{1,2}\to (\pm b,0)}\Bigg \{ \frac{32}{\sqrt{2}} E^2 G^2 \theta(x^+)\theta(x^-) \notag\\&
\times   \Bigg \{ \log \left(k|\vec{b}_2-\vec{b}_1|\right) \partial_{x^+} \left(  \frac{r_1}{r_1^2+2 (x^{\pm})^2} \theta(\tau-r_1) \right) \notag\\&
+\frac{1}{2 x^+}\Bigg[ \frac{b_{11}-b_{21}}{|\vec{b}_2-\vec{b}_1|^2}\theta(\tau-r_1)  \partial_{x^1}  \left(r_1 \frac{\tau^2-r_1^2}{r_1^2+2 (x^+)^2} \right) \notag\\&
+ \big(1 \leftrightarrow 2 \big) \Bigg]   \Bigg\}\Bigg\}  ,                         \label{g++2} \\
     g_{+-}^{(2)} &= \lim_{\vec{b}_{1,2}\to (\pm b,0)}\Bigg \{ 
     16 E^2 G^2 \theta(x^+)\theta(x^-) \text{sech} \hspace{0.02in} \eta   \notag\\& \times
     \Bigg\{\frac{1}{2 \tau}  \log \left(k|\vec{b}_2-\vec{b}_1|\right)\delta(\tau-r_1) \notag\\&
+\left[    \partial^2_{b_{11}b_{21}}-\frac{1}{4} \left(\frac{1}{\tau^2}\text{sech$^2$}\hspace{0.02in}\eta+\frac{1}{8}\tau \partial_{\tau}  \left(\frac{1}{\tau}\partial_{\tau}           \right) \right)    \right]  \notag\\& \times
{\cal J} (r_1,r_2,\tau) +\Big( 1\leftrightarrow2 \Big) \Bigg\} \Bigg\} , \label{+-2} \\
  g_{+1}^{(2)} &=\lim_{\vec{b}_{1,2}\to (\pm b,0)}\Bigg \{ 
  \frac{32}{\sqrt{2}} E^2 G^2 \theta(x^+)\theta(x^-)  \notag\\& \times
  \Bigg \{    \frac{b_{11}-b_{21}}{|\vec{b}_2-\vec{b}_1|^2}\frac{r_1}{r_1^2+2 (x^{\pm})^2} \theta(\tau-r_1)  +\frac{1}{2}  (\partial_{b_{21}})   \notag\\& \times
   \left [\frac{1}{1+e^{ \pm 2\eta}}\partial_{\tau} -\frac{1}{4 \sqrt{2} \tau} \text{sech}^2\hspace {0.02in}\eta  \right ] {\cal J} (r_1,r_2,\tau)    \Bigg\}  \Bigg\} , \label{g+12} \\
 g_{11}^{(2)} &=\lim_{\vec{b}_{1,2}\to (\pm b,0)}\Bigg \{ 
 -16 E^2 G^2 \theta(x^+)\theta(x^-) \text{sech} \hspace {0.02in}\eta \notag\\&
 \times \Big\{ \partial_{b_{11}b_{21}}^2 +\partial_{b_{11}b_{11}}^2 +  \partial_{b_{21}b_{21}}^2  \Big\}   {\cal J} (r_1,r_2,\tau)  \Bigg\}          ,                      \label{g112} \\
 g_{12}^{(2)} &=\lim_{\vec{b}_{1,2}\to (\pm b,0)}\Bigg \{ 
 -8 E^2 G^2 \theta(x^+)\theta(x^-) \text{sech}\hspace {0.02in}\eta\notag\\
\nonumber  &\Big\{ \partial_{b_{22}b_{11}}^2 +\partial_{b_{12}b_{21}}^2 +2\partial_{b_{11}b_{12}}^2  +2\partial_{b_{21}b_{22}}^2  \Big\}&\\   
& \times {\cal J} (r_1,r_2,\tau)          \Bigg\}   .&                             \label{g122} 
\end{align}
\end{subequations}
where
\begin{align}\label{taueta}
\tau=\sqrt {2 x^+ x^-} , \hspace {0.1in} \eta=\frac{1}{2}\log\left(\frac{x^+}{x^-}\right)  \hspace {0.05in} \mbox{and}  \hspace {0.05in} x^{\pm}=\frac{1}{\sqrt{2}}\tau e^{\pm \eta}. 
\end{align}
It is remarked that the variables $\tau$ and $\eta$ appearing in (\ref{gmn2}) should be thought as equal to their right hand side (see (\ref{taueta})) and not as a change of variables. 

\begin{figure}
\centering
\includegraphics[scale=0.55]{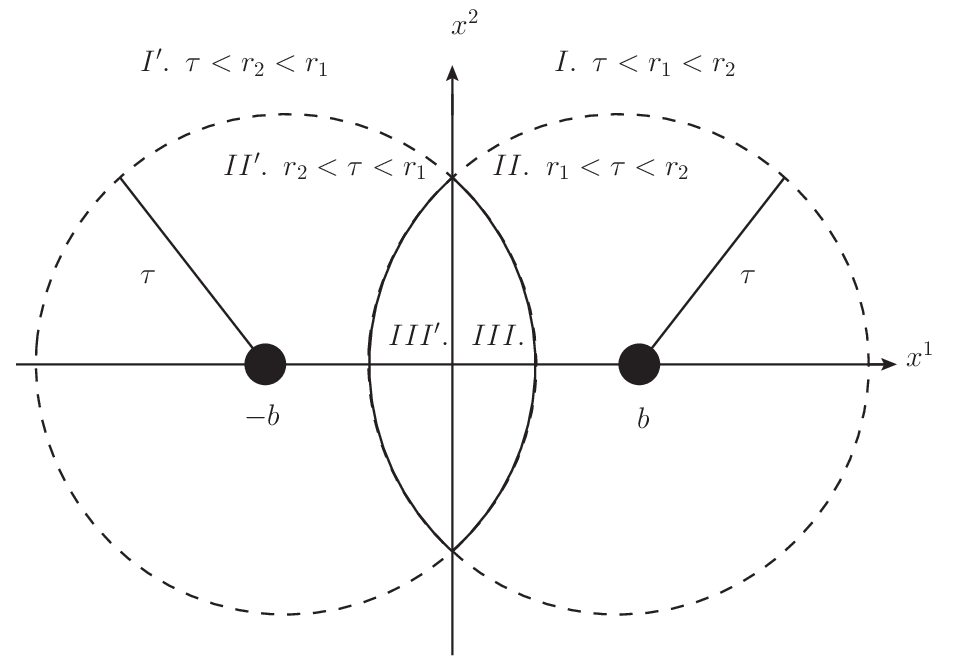}
\caption{The metric after the collision on the transverse plane. The centres of the shocks are located at $x^1=b$ and $x^1=-b.$ The two circles (dashed lines) have radius $\tau$.
At any given proper time $\tau$, the metric evolves differently in the three regions I, II and III (there is an obvious $\mathbb{Z}_2$ symmetry under $x_1 \leftrightarrow -x_1$ for the other three regions). The evolution is determined according to equations (\ref{J}), (\ref{Ja}) and (\ref{gmn2}).
Each region, determines whether the shocks have or have not enough proper time in order to propagate from the centers to the given region. For instance, region II defines the set of points where the shock with center at b has arrived but the shock with center at -b has not yet. Essentially, the evolution of the metric, according to this picture, is a manifestation of causality. In the $b=0$ limit there is a $\tau \leftrightarrow r$ symmetry.}
\label{re}
\end{figure}

The reason for introducing the vector $\vec{b}_{1,2}$ should now be obvious. Apart from the simplification of the calculation, one can obtain the remaining components $g_{\mu 2}^{(2)}$ from $g_{\mu 1}^{(2)}$ by interchanging $1 \leftrightarrow 2$, before taking the limits of equation (\ref{lim}).

It is also simple to obtain the $(-\mu)$ components from the $(+\mu)$ components by exchanging $+ \leftrightarrow -$ and $\vec{b}_1 \leftrightarrow \vec{b}_2$. We have now computed entirely the metric up to quadratic order in $E$ including back-reactions. This is the main result of our calculation.

In figure \ref{re}, we see a pictorial representation of the metric, described at the different regions, depending on the different values of the integral $J$ (see (\ref{J})).

\section{Bremsstrahlung Radiation}\label{br}

In section \ref{B2B} we have computed the corrections to the stress-energy tensor of the two massless particles. This provides us the necessary information in order to compute the bremsstrahlung radiation. As in \cite{Gal'tsov:2009zi}, one needs the polarization tensors and (the right hand side of) equations (\ref{deq}) in momentum space. Denoting with $k$ the 4-momentum, the spectral-angular distribution then reads
\begin{align}\label{log2}
\nonumber \frac{dE_{rad}}{d\omega d\Omega}&=\frac{G\omega^2}{2\pi^2} \sum_{pol}\Big| J_{\mu \nu}^{(2)}\epsilon^{\mu\nu}(k) \Big|^2\\
J_{\mu \nu}^{(2)} &\equiv T_{\mu \nu}^{(2)}(k)+S^{(2)}_{\mu \nu}(k)\, ,
\end{align}

where $\omega \equiv \frac{k^++k^-}{\sqrt{2}}$ is the frequency and $\epsilon^{\mu \nu} \left( k \right)$ are the graviton polarization tensors. The polarization tensors are derived in appendix \ref{B}.

\subsection{Estimating the radiated energy from dimensional analysis}
In order to guess the dependence of the radiated energy from the impact parameter and the energy, we can use simple dimensional analysis. 

The energy momentum tensor, $J_{\mu \nu}$, has dimensions of $\left[E\right]^4$, where $[E]$ denotes units of energy. Then the following is true
\begin{equation}
J^{\left(2\right)}_{\mu \nu}(x) \sim \left[E\right]^4 \sim E^2 G \frac{1}{\left[L\right]^4}.
\end{equation}
where $[L]$ implies dimensions of length, while E is the energy of the shock. In the second proportionality we have used that $J_{\mu \nu}^{(2)}$ is of second order in E and first order in G, as explicitly shown in (\ref{Tmnc}). The Fourier transformation of this quantity is
\begin{equation}
J^{\left(2\right)}_{\mu \nu}(k;b) \sim E^2 G f_{\mu \nu}\left(k;b\right) ,
\end{equation}
where $f_{\mu \nu}\left(k;b\right)$ is a set of dimensionless functions of $k$ and the impact parameter, $b$, which is the only remaining length scale.

As a result of (\ref{log2}) the radiated energy behaves like
\begin{equation}
E^{\left(2\right)}_{rad} \sim G \int d\omega  d\Omega \Big|\omega J_{\mu \nu}(k;b) \epsilon^{\mu \nu}(k)\Big|^2 _{|{\vec k}|=\omega}
\end{equation}
where the emitted radiation is taken on shell and thus must satisfy ${|{\vec k}|=\omega}$.
Taking into account that the polarization tensors are dimensionless and performing the integrations, the emitted energy is completely determined by dimensional analysis. In particular, in the absence of any IR cut-offs we must formally have
\begin{equation}
E^{\left(2\right)}_{rad}\sim G \int d\omega d\Omega   E^4 G^2 |\omega f  \left(k ; b \right)|^2 _{|{\vec k}|=\omega} \sim \frac{E^4 G^3}{b^3}.
\end{equation}
Therefore,
\begin{equation} \label{ErE}
\frac{E^{\left(2\right)}_{rad}}{E}\sim \frac{E^3 G^3}{b^3}.
\end{equation}
The question is whether the coefficient missing in (\ref{ErE}) is finite. In particular, we would like to address the question where any possible divergences come from and whether there is a way to regulate them. We argue that the underlying coefficient is not well-defined in the absence of appropriate cut-offs and we will attempt to give an explanation. 

The first step is to show that any rotationally symmetric physical shock-wave\footnote{Which we define as the shock created by any positive, integrable rotational symmetric distribution $\rho(pr)$ which generalizes the point-like $\delta^{(2)}(r)$ distribution of the point-particle.}, grows as $\log(pr)$ at large $r=\sqrt{(x_1)^2+(x_2)^2}$ where $p$ is some transverse scale. Indeed, the radial part $\phi(r)$ of $t_{1}$ and of $t_2$ satisfy $\nabla^2_{\perp}\phi(r) \sim \rho(pr)$ where $p$ is the transverse scale that fixes the width of $\rho$. Then, the slope of $\phi$ is $\phi' \sim 1/r\int_0^r r\rho dr\rightarrow 1/r$ as $r\rightarrow \infty$ because $\rho$ is integrable by assumption. Since $\phi' \sim 1/r$ at large $r$ it implies that $\phi \sim \log(pr)$.

The second step is to consider the quantity $J^{(2)}_{+-}$ from the right hand side of (\ref{+-}) and in particular the term $\sim t_{1,x^+} t_{2,x^-}$. According to the previous statement the transverse part of this term grows as $t_{1,x^+} t_{2,x^-} \sim \log^2(pr)$ as $r\rightarrow \infty$. This term corresponds precisely to the components of the Riemann tensor, components, $R_{\pm\pm\mp\pm}$, that diverge as $r \rightarrow \infty$ (see discussion in section \ref{IRc}) when the two metrics are superimposed. The rest of the terms in $J^{(2)}_{+-}$ decay as $r\rightarrow \infty$.

The third step is to compute the Fourier transformation of $J^{(2)}_{+-}$ in order to apply (\ref{log2}). Evidently, such a Fourier transformation is not well defined because it suffers from an IR divergence at large r.

We thus conclude that using the perturbative expansion $GE/b$ for such a geometrical configuration where the sources move with the speed of light in d=4 space-time dimensions is problematic when one attempts to compute the radiation of the any two gravitationally interacting sources. Since this statement applies for any physical transverse distribution, our conclusion is rather universal.

Relaxing one of the conditions could make the computation of the radiation feasible. For instance such a problematic behaviour for the Fourier transformation would not appear in higher dimensions\footnote{An attractive scenario would be a shock wave collision in the presence of extra dimensions along the lines of \cite{Taliotis:2012qg}.} because the shocks at large distances fall of as $1/r^{d-4}$. Likewise, there are no issues appearing for particles moving with finite speed as the one's considered in \cite{Gal'tsov:2009zi,Constantinou:2011ju,Gal'tsov:2010me,Gal'tsov:2012sn}. 

Another possibility to regulate the problem would be to put a sharp IR cut-off at some $r_{IR}=1/p$ along the lines of section \ref{IRc} and of equation (\ref{Ebr}). According to \cite{Weinberg:1965nx}, the combined IR divergences arising from the quantum mechanical radiative corrections and from the classical  Bremsstrahlung radiation are cancelled when a resumption procedure is performed. However, such a cancellation applies only for the collinear extremely soft photons and gravitons. It is thus unclear whether such a cancellation applies for our case. 

Another related series of works to ours is found in  \cite{Gal'tsov:2009zi,Constantinou:2011ju,Gal'tsov:2010me,Gal'tsov:2012sn} where the authors study the radiation of massive particles that collide with an impact parameter. Their set-up allows them to take the massless limit provided the impact parameter is simultaneously taken to infinity. In this case, they find that the total radiated energy is zero. As we will see, this result is consistent with the example studied in section \ref{grwa}. It is pointed that the analysis of  \cite{Gal'tsov:2009zi,Constantinou:2011ju,Gal'tsov:2010me,Gal'tsov:2012sn}
 in the massless limit and finite impact parameter is rather inconclusive.

Our conclusions, methods and applicability region could be compared with the ones derived in \cite{Coelho:2013zs,Coelho:2012sy,Coelho:2012sya,Herdeiro:2011ck} where the authors consider a different avenue in organizing their perturbation scheme. They assume a strong and a weak shock and they expand along the light cone where the strong shock is located.

\subsection{Example of radiation from gravitational waves}\label{grwa}

In this example, we consider the collision of gravitational waves which, by definition, correspond to a zero $T_{\mu \nu}$. For simplicity we consider homogeneous waves in the transverse direction\footnote{One could argue that such shocks are a pure gauge. Indeed, when there is a single shock moving (say) along $x^+$, the transformation $x^+ \rightarrow x^+ +1/2 \int t_1(x^-) dx^-$ removes the $t_1(x^-)(dx^-)^2$ component of the metric. However, when two shocks are superimposed such a transformation would not work in the forward light cone because there are mixing terms.}. It thus makes sense to compute the $E_{rad}/V_{x^1x^2}$ where $V_{x^1x^2}$ is the transverse volume. We expect that the total radiation will be zero as we collide zero energy shocks. 

The only non trivial component of (\ref{deq}) is the (\ref{+-}) component where only the last term in the right hand side remains. The second order solution is $g_{+-}=-1/4 t_1(x^+)t_2(x^-)$ while $J^{(2)}_{+-}=1/2t_{1,x^+} t_{2,x^-}$. The Fourier transformation of $J^{(2)}_{+-}(x^{\mu})$ is $J^{(2)}_{+-}(k) \sim \delta^{(2)}(\bf{k}_{\perp})$ where the proportionality constant depends on the details of the profile of $t_1(x^{+})$ and $t_2(x^{-})$. 

The last step is to use (\ref{log2}) by contracting $J^{(2)}_{+-}(k) $ with $\epsilon^{+-}_{(i)}\,=-\frac {k_{\perp}^2} {2\sqrt{2}(k_{\perp}^2+1/2(k_+-k_-)^2)}$ for $i=I,\,II$ (see appendix \ref{B}) which, when the graviton is on shell, yields $\epsilon^{+-}_{(i)} \,= -\frac {k_{\perp}^2} {\sqrt{2}(k_++k_-)^2}$.
Combining all the previous information and using (\ref{log2}) we finally obtain
\begin{align}\label{0E}
E_{rad}/V_{x^1x^2} \sim E_{rad}/\delta^{(2)}(0) \sim \int dk_{\perp} k_{\perp}\frac {k_{\perp}^4 \delta^{(2)}(\bf{k}_{\perp}) } {(k_++k_-)^4}=0
\end{align}
where we used that the transverse volume is proportional to $\delta^{(2)}(0)$. Thus, we find that the total energy per transverse area radiated from gravitational waves is zero as expected.

In fact, it is not hard to argue that this is an all order result and hence an exact statement. The reason is that, since $J_{\mu \nu}$ in (\ref{log2}) has no transverse dependence, its Fourier transformation will be $\sim \delta^{(2)}(\bf{k}_{\perp})$. Contracting then $J_{\mu \nu}(k)$ with any of the polarization tensors of (\ref{pol1}) and (\ref{pol2}) would yield zero because all the components are proportional to either $k_1$ or to $k_2$.

In order to make contact with \cite{Gal'tsov:2009zi,Constantinou:2011ju,Gal'tsov:2010me,Gal'tsov:2012sn}, one must consider the massless limit and in addition take the impact parameter to infinity. In this case, as already mentioned in last section, works \cite{Gal'tsov:2009zi,Constantinou:2011ju,Gal'tsov:2010me,Gal'tsov:2012sn} yield a zero radiative energy. Likewise, if we start from the shocks (\ref{s12}) and take the $b\rightarrow \infty$ limit simultaneously with $p\rightarrow 0$ such that $bp=$fixed it implies that we collide the massless particles very far from each other while the IR cut-off is taken to zero\footnote{The same occurs in \cite{Gal'tsov:2009zi,Constantinou:2011ju,Gal'tsov:2010me,Gal'tsov:2012sn} where there is not an IR cut-off.}.
 The resulting shocks when these two limits are taken correspond precisely into two transversally homogeneous gravitational waves with zero $T_{++}$ and $T_{--}$. According to (\ref{0E}), these two waves radiate zero energy; exactly as in \cite{Gal'tsov:2009zi,Constantinou:2011ju,Gal'tsov:2010me,Gal'tsov:2012sn}.
 

\section{Conclusions}\label{scon}

We have studied the causal, purely gravitational, collision of two massless shockwaves, having a non-zero impact parameter taking into account back reactions following a perturbative treatment. Our main conclusions as summarized as follows.

\begin{enumerate}

\item Our main computational result is the derivation of the second order corrections to the metric in the presence of an impact parameter $b$, taking into account the back-reaction. This has been presented in equation (\ref{gmn2}) and pictorially in fig. \ref{re}. Fig. \ref{re} describes intuitively the manner which the metric evolves in time and how this evolution is in harmony with causality as it would be expected. In the $b=0$ limit, equations (\ref{J}), (\ref{Ja}) and fig. \ref{re} suggest a $\tau \leftrightarrow r$ symmetry. A similar symmetry has been observed in heavy ions in \cite{Taliotis:2010pi} and in \cite{Gubser:2010ze} using different approaches. 

\item In fact, it seems that the evolution of space-time soon after the collision is qualitatively similar as the situation with the expanding plasma in heavy ions in the following sense. During the first stages of the collision the plasma is thin in the longitudinal direction and due to the larger pressure, it has the tendency to expand and isotropize \cite{Kovchegov:2005ss,Kovchegov:2005az}. Likewise, in the present set-up which, according to sec. \ref{IRc} is an early times approximation, the metric is localized in the vicinity of $x^+\sim x^- \sim 0$ due to the $\theta$ functions in equation (\ref{J}). Another way to see the localization of the metric along the collision direction is from a trapped surface analysis perspective where the trapped surface at $t=0^+$ is 2-dimensional \cite{Eardley:2002re,Taliotis:2012sx}. On the other hand, we expect that for sufficiently large energy, the final product will be a (spherically symmetric) Schwarzschild black hole showing that the produced metric will eventually isotropize, just like the produced medium in heavy ions.

\item For zero impact parameter, the perturbative treatment and consequently our approximation breaks down, since our expansion parameter, $\frac{E G}{b}$, diverges. When the impact parameter $b$ is zero, the energy momentum tensor is no longer conserved and thus the Bianchi identities are also violated. We believe that this is a sign of non-perturbative effects that start to become dominant and that a perturbative treatment is no longer justified in this regime. It is  pointed out that such a violation is point-like in space and instantaneous in time and is hidden behind a horizon that forms after the collision (see \cite{Eardley:2002re}).

\item We have shown that the trace of the energy momentum tensor is zero, not only up to first order (negative times), but also for up to second order (after the collision). This indicates that the energy momentum tensor could be traceless up to all orders, suggesting a sort of conservation of tracelessness: one starts with a traceless energy momentum tenor at some initial time and this tracelessness continues to apply all the way during the time evolution. It is an interesting speculation which, certainly should be given further attention. 

\item Our approach is perturbative and the region where it is valid is discussed in section \ref{IRc}. The discussion in that section extends beyond the analytical predictability of our perturbative approach. In particular, we guess that the equilibration time for a black hole to be formed at high energies must be given by $t_{eq} \sim EG$ which is the only available scale either in the $b=0$ or in the $b\ll EG$ limit. This might seem counter-intuitive in the sense that in high energies one would expect things to be developed faster: for instance in \cite{East:2012mb}, increasing the boost factor results in decreasing $t_{eq}$\footnote{We thank F. Pretorious for a private communication on this issue and for discussing \cite{East:2012mb}. The set up used in \cite{East:2012mb} involves the gravitational collision coupled to a perfect fluid rather to point-like particles.}. We argue that the set-up we have is different in the sense that the boost factor in our case is always infinite. We argue that by increasing the energy more, the process becomes more violent and it takes more time to equilibrate. Definitely, the numerical relativity approach which, unfortunately has limitations in taking large values of the boost-factor, is the most reliable avenue to explore such a question. 

\item The estimation of the the gravitationally radiated energy with respect to the energy, $E$, and the impact parameter, $b$ is discussed. As noted, positive energy transverse distributions create shocks with a universal behaviour: at large distances from the center of these sources, the shocks  grow logarithmically instead of decaying. This implies that such shocks interact strongly for an infinite portion of space and this, in the present approach, would produce infinite radiation. We propose ways to regulate this issue and we show that this is a fact of fast moving, everywhere positive definite, transverse distributions in d=4 space-time dimensions. By relaxing one of these conditions, for example  considering zero energy gravitational waves, we are able to compute the radiation. In this example we find that the total radiation produced is zero. We argued that such a result applies to all orders as it would be expected by conservation considerations.

\item{
We showed in appendix \ref{A} that for dilute enough transverse distributions of the energy, a black hole can not be formed during head-on collisions. Equivalently, a black hole is formed
when transversally-extended distributions are collided head-on only if the collision energy is sufficiently large compared to the width of the distributions. In other words, dense enough distributions are required in order to form a black hole and our analysis makes quantifies this statement}

\end{enumerate}

\acknowledgments

We would like to thank B. Craps, U. Gerlach, S. Giddings, S. Gubser, G. Horowitz, I. Iatrakis, Y. Kovchegov, S. Mathur, F. Pretorious, N. Toumbas and especially C. Herdeiro, M. Sampaio and Th. Tomaras for reading the manuscript and for informative discussions. In addition, A.T. wishes to thank the CCTP at The University of Crete for creating a stimulating atmosphere to initiate this work.


A.T. is supported in part by the Belgian Federal Science Policy Office through the Interuniversity Attraction Pole P7/37, and in part by the ÒFWO-VlaanderenÓ through the project G.0114.10N, which he acknowledges both.

This work (including A.T.) was (in addition) supported in part by grants PERG07-GA-2010-268246, PIF-GA-2011-300984, the EU program ``Thales'' and "HERAKLEITOS II'' ESF/NSRF 2007-2013 and was also co-financed by the European Union (European Social Fund, ESF) and Greek national funds through the Operational Program ``Education and Lifelong Learning'' of the National Strategic Reference Framework (NSRF) under ``Funding of proposals that have received a positive evaluation in the 3rd and 4th Call of ERC Grant Schemes''. 


\appendix

\renewcommand{\theequation}{A\arabic{equation}}
  \setcounter{equation}{0}
\section{Trapped Surfaces from Extended Sources}
\label{A}

In this appendix, we study the behaviour of shockwave collisions arising from extended sources on the transverse plane. This investigation has been carried out in great detail in \cite{Taliotis:2012sx} (see also \cite{Kohlprath:2002yh}) where the author has generalized the investigation to any extended sources satisfying reasonably physical and quite mild constrains and for both, flat and $AdS$ backgrounds\footnote{For AdS backgrounds on trapped surfaces there have been many interesting works including \cite{Gubser:2008pc,Kovchegov:2009du,Arefeva:2012ar,DuenasVidal:2012sa}}.

Here, using a simple model, we show that when extended sources on the transverse plane are collided at zero impact parameter, two trapped surfaces (a small and a large one) are obtained for sufficiently large energy\footnote{ We thank G. Veneziano for a relevant discussion.}.

In order to see this, we replace the delta function localized matter distribution with an extended one and in particular we consider a transversally symmetric distribution $\rho$. Hence, we assume a  $T_{++} \sim E \rho(kx_{\perp}) \delta(x^+)$ where $x_{\perp}^2=(x^1)^2+(x^2)^2$ and where $k$ fixes the width of $\rho$ which can be energy ($E$) depended or energy independent. The stress-tensor can be normalized such that when $T_{++}$ is integrated to yield $E$. The corresponding transverse part, denoted by $\phi$, of the shock satisfies $\nabla_{\perp}^2 \phi=\rho$. This yields
\begin{align}\label{ph}
\phi \sim E G \int_0^r dr \left(\frac{\int_0^{r'} dr'' r'' \rho(r'')}{r'}\right).
\end{align}

The trapped surface consists of two pieces, $S_{+}$ and $S_-$. These are parametrized with the help of two functions, $\psi_+$ and $\psi_-$\footnote{For details we refer the reader to the appendices of \cite{Kiritsis:2011yn,Taliotis:2012qg}.} which satisfy the following differential equation
\begin{align}\label{de}
 \nabla^2_{\perp} (\psi_{\pm}-\phi_{\pm})=0.
\end{align}

It is pointed out that $\nabla^2_{\perp}  \phi_{\pm}$ provides a source term for  $\nabla^2_{\perp}  \psi_{\pm}$. The missing ingredient is the boundary conditions. Dropping the indices $\pm$ from $\psi_{\pm}$ from now on assuming a zero impact parameter and identical shocks we have $\psi_{+}=\psi_{-}=\psi$. The boundary conditions then read
 \begin{align}\label{BC}
\psi \Big |_C=0 \hspace{0.4in}  \sum_{i=1,2}\left[ \nabla_{x^i} \psi  \nabla_{x^i} \psi  \right]  \Big |_{C}=8
 \end{align}
 for some curve $C$ which defines the boundary of the trapped surface and where both, $S_+=S$ and $S_-=S$ end.

The function $\psi$ satisfying the left boundary condition in (\ref{BC}) is given by $\psi(x_{\perp})=\phi(x_{\perp})-\phi(x^c_{\perp})$ which vanishes on the trapped surface defined by $x_{\perp}=x^c_{\perp}$. Imposing the right condition, allows one to specify $x^c_{\perp}$ in terms of the parameters of the problem, namely $EG$ and $k$. Thus from $(\psi'(x_{\perp}))^2 \sim 1$ one obtains
 \begin{align}\label{ps1}
&(\psi'(x^c_{\perp}))^2=(\phi'(x^c_{\perp}))^2 \sim 1 \notag\\&
=>\frac{\int_0^{y_c} dr y \rho(y)}{y_c} \sim \frac{1}{EkG}, \,\, y\equiv k x_{\perp}. 
\end{align}
The last equation provides the condition of a trapped surface from colliding extended sources: for a given energy $E$ and a transverse width $k$, a trapped surface exists if there is a $y_c$ such that (\ref{ps1}) has a solution. Taking for concreteness a Gaussian distribution $\rho=k^2 e^{-x_{\perp}^2k^2}$ equation (\ref{ps1}) yields
\begin{align}\label{ps}
\left(  \frac{e^{-k^2 x_{\perp}^2}-1}{kx_{\perp}}\right)^2\sim \left(\frac{1}{Ek G}\right)^2
\Big|_{x_{\perp}=x^c_{\perp}}.
\end{align}

A few remarks are in order.
\begin{itemize}
\item{The function $\left((e^{-y^2}-1)/y\right)^2$ for positive $y$ becomes zero at $y=0$ and $y=\infty$ and is strictly positive with a maximum at $y \approx 1.12$ (see figure \ref{TSd}).}

\begin{figure}
\includegraphics[scale=0.62]{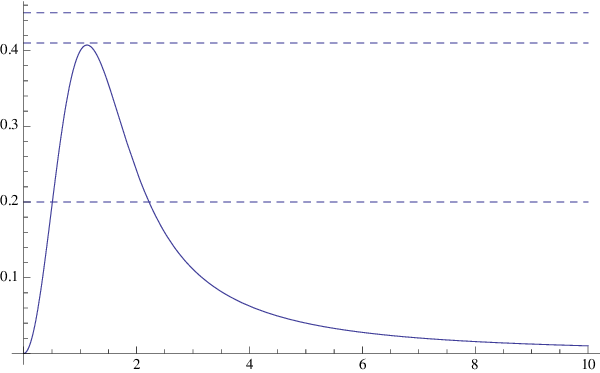}
\caption
 {\label{TSd}
  The function $\left((e^{-y^2}-1)/y\right)^2$ as a function of $y\equiv k x_{\perp}$.}
\end{figure} 

\item{The previous statement implies that for a given sufficiently large energy $E$ there is a small trapped horizon and a large trapped horizon. In addition, according to figure \ref{TSd}, there is a critical value of the quantity $EkG$ such that the small and the large apparent horizons merge. Finally, for smaller values of the quantity $EkG$, the trapped surface can not exist  (see top dashed line in the figure). This implies that for small energies or for large widths (small $k$'s; very dilute matter) the trapped surface can not be formed. This is an expected result.}

\item{The full classification of all the distributions $\rho$ and the kind of surfaces that they create has been performed in \cite{Taliotis:2012sx}. Here we briefly mention the basic features for completeness. There are three classes of trapped surfaces: (a) The ones created for  energy no matter how small it is ($k$ is assumed fixed). (b) Trapped surfaces with a single trapped horizon which are created only for sufficiently large energies. (c) Trapped surfaces with two trapped horizons which are created only for sufficiently large energies (see for example fig. \ref{TSd}). In all the cases, the shocks grow logarithmically at infinity while in the high energy limit $EkG\gg1$, the entropy grows as $S_{trap} \sim E^2G$. Such a growth applies for any distribution $\rho$ and hence the result is universal.}

\end{itemize}

\renewcommand{\theequation}{B\arabic{equation}}
\section{Polarization tensors}
\label{B}
We now proceed to derive the polarization tensors. There are two such tensors, $\epsilon^{\mu \nu}_{(1)}$ and $\epsilon^{\mu \nu}_{(2)}$. It is necessary to  define two space-like unit vectors, $e^M_1$ and $e^M_2$, that are both perpendicular among themselves and to the wave vector of the radiated gravitational radiation, $k^\mu$, before one can derive the polarization tensors. These vectors satisfy the following relations
\begin{equation}
e_\alpha^\mu e_{\beta \mu}=\delta_{\alpha \beta}, \, \, e_\alpha^\mu k_\mu=0\, .
\end{equation}
The two vectors explicitly written in light-cone coordinates read
\begin{align}
\nonumber
e_1^\mu&=\left(-\frac{k_{\perp}}{\sqrt{2} |\vec{k}|},\frac{k_{\perp}}{\sqrt{2} |\vec{k}|},\frac{k_1 \left(k_+-k_-\right)}{\sqrt{2} k_{\perp}|\vec{k}|},\frac{k_2 \left(k_+-k_-\right)}{\sqrt{2} k_{\perp}|\vec{k}|}\right)\\
e_2^\mu&=\left(0,0,-\frac{k_2}{k_{\perp}},\frac{k_1}{k_{\perp}}\right)\, ,
\end{align}
where we have defined $\vec{k}\equiv\left(k_1,k_2,\frac{k_+-k_-}{\sqrt{2}}\right)$.

We can now proceed to construct the two polarization tensors. They should, by construction, be orthogonal to each other and traceless, i.e. satisfy the following relations
\begin{equation}
\eta_{\mu \nu}\epsilon_{a}^{\mu \nu}=0, \, \epsilon_a^{\mu \nu} \epsilon_{b \mu \nu}=\delta_{a b}
\end{equation}
One can easily see that the two polarization tensors can be written in terms of the polarization vectors as,
\begin{equation}
\epsilon_{(I)}^{\mu \nu}=\frac{e_1^\mu e_1^\nu+e_2^\mu e_2^\nu}{\sqrt{2}}\, , \epsilon_{(II)}^{\mu \nu}=\frac{e_1^\mu e_1^\nu-e_2^\mu e_2^\nu}{\sqrt{2}}\, .
\end{equation}
\\
Writing the explicit form of the two tensors, we have
\begin{tiny}
\begin{align}\label{pol1}
\nonumber 
&\epsilon_{(I)}^{\mu \nu} =\frac{1}{2\sqrt{2}|\vec{k}|^2} \times \\
& \begin{pmatrix}
       k_{\perp}^2 & -k_{\perp}^2 & k_1\left(k_--k_+\right) & k_2\left(k_--k_+\right)           \\[0.2em]
       -k_{\perp}^2 & k_{\perp}^2           & k_1\left(k_+-k_-\right) & k_2\left(k_+-k_-\right) \\[0.2em]
       k_1\left(k_--k_+\right)           & k_1\left(k_+-k_-\right) & 2k_2^2+\left(k_+-k_-\right)^2 &-2 k_1k_2 \\[0.2em]
		k_2\left(k_--k_+\right) &     k_2\left(k_+-k_-\right) &-2 k_1k_2 & 2k_1^2+\left(k_+-k_-\right)^2
     \end{pmatrix},
     \end{align}
     \end{tiny}
     
     \vspace{-0.5in}

\begin{tiny}
\begin{align}\label{pol2}
\nonumber
&\epsilon_{(II)}^{\mu \nu} =\frac{1}{2\sqrt{2}|\vec{k}|^2} \times  \\ 
&\begin{pmatrix}
       k_{\perp}^2 & -k_{\perp}^2 & k_1\left(k_--k_+\right) & k_2\left(k_--k_+\right)           \\[0.1em]
       -k_{\perp}^2 & k_{\perp}^2           & k_1\left(k+-k_-\right) & k_2\left(k_+-k_-\right) \\[0.1em]
       k_1\left(k_--k_+\right)           & k_1\left(k+-k_-\right) & -\frac{2k_2^2|\vec{k}|^2+k_1^2\left(k_+-k_-\right)^2}{k_{\perp}^2} &\frac{2 k_1 k_2 \left(k_{\perp}^2+\left(k_+-k_-\right)^2\right)}{k_{\perp}^2} \\[0.1em]
		k_2\left(k_--k_+\right) &     k_2\left(k_+-k_-\right) &\frac{2 k_1 k_2 \left(k_{\perp}^2+\left(k_+-k_-\right)^2\right)}{k_{\perp}^2}  & -\frac{2k_1^2|\vec{k}|^2+k_2^2\left(k_+-k_-\right)^2}{k_{\perp}^2}
     \end{pmatrix}.
     \end{align}
     \end{tiny}



\bibliography{references}
\bibliographystyle{unsrt}
\end{document}

\end{document}